\documentclass[a4paper,12pt]{article}
\usepackage{amsmath,amssymb,amsthm,mathrsfs,graphicx,float,colordvi,url,wrapfig,color} 
\usepackage{enumerate}
\usepackage{enumitem} 

\def\cG{{\cal G}}  
\def\tm{\tilde{m}}   
\def\tD{\tilde{\Delta}} 

\begin{document}
\renewcommand{\thefootnote}{\fnsymbol{footnote}} 
\begin{titlepage}

\vspace*{0.0mm} 

\begin{center}
{\large \bf 
Kerr/CFT correspondence in a 4D extremal rotating \\ regular
\vspace{1.5mm}
black hole with a non-linear magnetic monopole
}


\vspace*{8mm}

\normalsize
{\large {Shingo Takeuchi\footnote{shingo.portable(at)gmail.com}}} 

\vspace*{6mm} 

\textit{
The Institute for Fundamental Study, ``The Tah Poe Academia Institute''},\\
\vspace*{0.10 cm}
\textit{Naresuan University Phitsanulok 65000, Thailand}\\

\end{center}


\vspace*{5mm}

\begin{abstract}
We carry out the Kerr/CFT correspondence in a four-dimensional extremal rotating regular black hole with a non-linear magnetic monopole~(NLMM).    
One problem in this study would be whether our geometry can be a solution or not.  
We search for the way making our rotating geometry into a solution based on the fact that the Schwarzschild regular black hole geometry with a NLMM can be a solution.  
However, in the attempt to extend the Schwarzschild case that we can naturally consider, it turns out that it is impossible to construct a model in which our geometry can be a exact solution.
We manage this problem by making use of the fact that our geometry can be a solution approximately in the whole space-time except for the black hole's core region. 
As a next problem, it turns out that the equation to obtain the horizon radii is given by a fifth-order equation due to the regularization effect.  
We overcome this problem by treating the regularization effect perturbatively. 
As a result, we can obtain the near-horizon extremal Kerr (NHEK) geometry with the correction of the regularization effect.    
Once obtaining the NHEK geometry, we can obtain the central charge and the Frolov-Thorne temperature in the dual CFT.  
Using these, we compute its entropy through the Cardy formula, which agrees with the one computed from the Bekenstein-Hawking entropy.

\end{abstract}

\end{titlepage}

\newpage  

\section{Introduction} 

Getting understanding for the microscopic states of the Bekenstein-Hawking entropies is one of the very important issue for us.  
Going through theoretical developments such as 
the holographic principle~\cite{tHooft:1993dmi,Susskind:1994vu,Bigatti:1999dp},  
the discovery of the D-branes~\cite{Polchinski:1995mt}, 
the so-called Strominger-Vafa~\cite{Strominger:1996sh}, 
and the AdS/CFT~\cite{Maldacena:1997re,Gubser:1998bc,Witten:1998qj},       
at 2008, in a four-dimensional extremal Kerr black hole, \cite{Guica:2008mu} could succeed in 
reading off the central charge and the temperature in the dual CFT    
which can leads to the entropy exactly agreeing with the Bekenstein-Hawking entropy.    
This computation is considered as the microscopic computation of the Bekenstein-Hawking entropy, which is refereed to as the Kerr/CFT correspondence.

Thanks to the Kerr/CFT correspondence, we have now reached the stage that 
the Bekenstein-Hawking entropy in all the classical Kerr black holes at the extremal limit can be exactly reproduced in the dual 2D CFT\footnote{
E.g.~(i)~For arbitrary dimensions and extension to gauged/ungauged SUGRA,~\cite{Lu:2008jk,Chow:2008dp}, 
(ii)~for general expression of conserved charges~\cite{Hartman:2008pb, Compere:2009dp},
(iii)~for Kerr-Newman BH with arbitrary cosmological constant,~\cite{Hartman:2008pb}, 
(iv)~for 5D Kaluza-Klein BH,~\cite{Azeyanagi:2008kb},  
(v)~for 5D black rings,~\cite{Loran:2008mm, Goldstein:2011jh, Chen:2012yd},  
(vi)~for 5D D1-D5-P and BMPV BH's,~\cite{Isono:2008kx, Azeyanagi:2008dk}, 
(vii)~for higher order derivative corrections,~\cite{Krishnan:2009tj, Azeyanagi:2009wf, Goldstein:2011jh, Hayashi:2011uf}, 
(viii)~for Kerr/CFT based on M/superstring theories,~\cite{Guica:2010ej,Compere:2010uk}}. 
Hence our next issue would be to extend Kerr/CFT correspondence to more actual one. 
One way in this attempt is the extension to the non-extremal~(e.g.~\cite{Castro:2010fd,Chen:2010yu,Chen:2010as,Chen:2010xu,Chen:2010ywa,Wang:2010qv,Chen:2011kt,Chen:2010jc,Krishnan:2010pv,Matsuo:2009sj}).    
Another one would be to consider some more actual black holes.

In such a situation, we will apply the Kerr/CFT correspondence to a four-dimensional regular rotating extremal black hole.   
We explain what the regular black holes are in what follows. 
\newline

The core of classical black holes is singular.  
However it is expected that quantum gravity effects working among the matters in the core would become strong repulsive force as those matters close up each other to around the plank length. 
Therefore there is an idea that indeed the singularity is not present in actual black holes. 
As a result we come to consider the black hole models without singularities, which we name {\it regular black hole} or {\it non-singular black hole}.  
The regular black holes would be correspond to the actually observed black holes~(e.g. GRS 1915+105~\cite{McClintock:2006xd} and cygnus X-1~\cite{Gou:2011nq}).

The key concept in the formulation of the regular black holes is to suppose that 
with remaining the asymptotic region of the black hole space-times as the classical ones, only the core region is given as a de Sitter space 
by considering that the attractive force among the matters in the core is balanced with the expanding behavior of de Sitter spaces.
The regular black holes proposed so far can be categorized into the three types:~(i)~given with a non-liner electron/magnetic monopole 
fields~\cite{Bardeen1968,Dymnikova:1992ux,Bronnikov:2000vy, AyonBeato:1998ub, AyonBeato:1999rg, AyonBeato:2000zs, AyonBeato:2004ih, Moreno:2002gg,Hayward:2005gi}~({\it first type}), 
(ii) given by connecting two space-times with a thin-shell~\cite{Sakharov:1966aja,thinshell1,thinshell2,Uchikata:2012zs,Uchikata:2015xma}~({\it second type}), 
or (iii) given with matters distributed according to the Gaussian distribution based on the non-commutative space-time conjecture~\cite{Nicolini:2005vd, Modesto:2010rv}~({\it third type}).

The regular black holes belonging to the first type have been firstly proposed by Bardeen~\cite{Bardeen1968}. 
The energy-momentum tensors leading the regular black holes into a solution have been investigated in \cite{Dymnikova:1992ux}.   
The regular black holes as a solution have been proposed~\cite{Bronnikov:2000vy, AyonBeato:1998ub, AyonBeato:1999rg, AyonBeato:2000zs, AyonBeato:2004ih, Moreno:2002gg, Balart:2014cga, Fan:2016hvf}.  
Formation and evaporation of the regular black holes have been investigated by Hayward~\cite{Hayward:2005gi}. 
The metric used in \cite{Hayward:2005gi} is called {\it Hayward type}. 
Rotating versions of the Bardeen and Hayward type regular black holes have been obtained in \cite{Bambi:2013ufa} using the Newman-Janis algorithm~\cite{Newman:1965tw,Newman:1965my}. 
On the other hand, the regular black holes in the second type are the ones 
formulated by connecting de Sitter and black hole space-times by putting a thin shell between those two~\cite{Sakharov:1966aja,thinshell1,thinshell2,Uchikata:2012zs,Uchikata:2015xma}.  
Finally, the regular black holes in the third type are the ones formulated by supposing that the matters of the gravity source are distributed according to the Gaussian distribution 
based on the non-commutative space-time conjecture~\cite{Smailagic:2003yb, Nicolini:2005vd, Modesto:2010rv}~({\it non-commutative type}).
\newline

Among the three types, the non-commutative type looks most realistic.  
However, its Lagrangian is unclear, and correspondingly whether the geometry is a solution or not is unclear. 
On the other hand, in the Schwarzschild Hayward type, by considering a non-liner magnetic monopole~(NLMM) or electric charge, the geometry can be a solution.  
Therefore, we will consider the rotating Hayward type geometry with a NLMM to be obtained from a natural extending the Schwarzschild type with a NLMM. 
We mention the organization of this paper.

In Sec.\ref{Chap:RBH}, the regular black hole geometry with a NLMM in this study is given, 
and in Sec.\ref{Chap:Sol}, considering an Einstein-nonlinear Maxwell action, we search for the way to make our rotating geometry with a NLMM into a solution. 
Then it turns that it is adding a new term vanishing at zero-rotation to the action, because the form of the geometry seems to have no space where we can add further modification for making into a solution.   
This would be the natural extension we can consider from the Schwarzschild case with a NLMM~\cite{Fan:2016hvf}. 
However, it has turned out that we cannot obtain the action exactly containing the rotation effect of our geometry as a solution. 
However, it turns out that it can satisfy the Einstein equation approximately in the whole space-time except for the black hole's core region. 
Making use of this fact, we manage this problem.  

Then, it turns out that the equation to obtain the horizon radii is  a fifth-order equation due to the regularization effect for the black hole's singularity.  
In order to overcome this, in Sec.\ref{Chap:NHEK}, we propose to treat the regularization effect perturbatively. 
By this, we can obtain the horizon radii and the near-horizon extremal Kerr~(NHEK) geometry of our geometry, perturbatively.

Once we obtain the NHEK geometry, we obtain the central charge and the temperature in the dual CFT in Sec.\ref{Chap:CC} and \ref{Chap:CFTT}.  
In Sec.\ref{Chap:EdCFT}, using these, we compute the entropy in the dual CFT through the Cardy.
In Sec.\ref{Chap:Summary}, we summarize this study.

In Appendix.\ref{App:CC} and \ref{App:CFTT}, a computation of the central charge based on the Lagrangian formalism~\cite{Barnich:2001jy, Barnich:2007bf}, 
and the expression of the Hawking temperature in our geometry are shown.

\section{Rotating regular black hole with a NLMM in this study}
\label{Chap:RBH} 

The rotating Hayward type regular black hole geometry with a NLMM we consider in this paper is
\begin{eqnarray}\label{RRBHTD}
ds^2 = g_{tt} dt^2 + 2 g_{t\phi} dt d\phi + g_{\phi\phi}d\phi^2 + g_{rr}dr^2 + g_{\theta\theta}d\theta^2 
\end{eqnarray}
with
\begin{eqnarray}\label{RR metric2}
g_{\mu\nu}=
\begin{pmatrix} 
-\frac{\tD - a^2 \sin^2\theta}{\Sigma}                     & 0                  & 0      & -\left(\frac{r^2+a^2 - \tD}{\Sigma} \right) a\sin^2 \theta \\  
0                                                          & \frac{\Sigma}{\tD} & 0      & 0 \\ 
0                                                          & 0                  & \Sigma & 0 \\  
-\left(\frac{r^2+a^2 - \tD}{\Sigma} \right) a\sin^2 \theta & 0                  & 0      &  \left( (r^2 + a^2)^2 - \tD a^2 \sin^2 \theta \right) \frac{\sin^2 \theta}{\Sigma}
\end{pmatrix},
\end{eqnarray} 
and a gauge field,
\begin{eqnarray}\label{RRBHTDG}
A=-\frac{q_m \cos \theta}{\Sigma} \big(a dt - (r^2+a^2) d\phi \big),
\end{eqnarray}
where $q_m$ is the charge of the NLMM, $\mu,\,\nu = t, r, \theta, \phi$,  and
\begin{eqnarray}
\label{SigmatD}
\Sigma &=& r^2 + a^2 \cos^2 \theta,\\
\label{tD}
\tD    &=& r^2 -2 \tm r +a^2 + q_m^2,
\end{eqnarray} 
where
\begin{eqnarray}\label{MinRBH}
\tm = \frac{m r^3}{r^3 + l_p^3}, \qquad a = \frac{L}{m}. 
\end{eqnarray}
$m$ is the mass of the black hole and $L$ is the angular momentum of the black hole,  
and $l_p \sim \textrm{the Plank length}$, which originates in the NLMM charge as mentioned in the next section.

\section{On what our geometry can become a solution}
\label{Chap:Sol} 

In this section, we consider an action enable to contain our geometry with a NLMM given in Sec.\ref{Chap:RBH} as a solution, and show how such a geometry can satisfy the Einstein equation. 
\newline

We start with the following regular Schwarzschild black hole geometry with a NLMM:
\begin{eqnarray}\label{SchRegBH}
ds^2=-fdt^2 + \frac{dr^2}{f}+r^2d\Omega^2, \quad A=Q_m \cos \theta d\phi
\end{eqnarray}
where
\begin{eqnarray}
f = 1 - \frac{2\alpha^{-1}q_m^3r^{2}}{r^3+q_m^3},
\end{eqnarray}
with $Q_m=q_m^2/\sqrt{2\alpha}$ and $2\alpha^{-1}q_m^3=m$. This can be an exact solution in the following model:~\cite{Fan:2016hvf} 
\begin{eqnarray}
I=\frac{1}{16\pi}\int d^4x\sqrt{-g}\left(R-\mathcal{L}\right) \quad {\rm with} \quad {\cal L}=\frac{12}{\alpha} \frac{(\alpha {\cal F})^{\frac{3}{2}}}{\left(1+(\alpha {\cal F})^{3/4}\right)^2},
\end{eqnarray}
where ${\cal F}=F_{\mu\nu}F^{\mu\nu}$. 
Actually, we can confirm this by the Einstein equation derived from the above action, 
\begin{eqnarray}
R_{\mu\nu}-g_{\mu\nu} R = -\frac{1}{2}g_{\mu\nu}{\cal L} -2 \frac{\partial \cal L}{\partial \cal F} F_\mu{}^\alpha F_{\alpha\nu}. 
\end{eqnarray}
\newline

If we attempt to obtain a regular rotating black hole as a solution, the geometry part would be obtained from eq.(\ref{SchRegBH}) using the Newman-Janis algorithm~\cite{Newman:1965tw,Newman:1965my}, 
which leads to eq.(\ref{RRBHTD})~\cite{Bambi:2013ufa}.   
Although the gauge filed part in eq.(\ref{SchRegBH}) should also be converted, currently there is no prescription for this. 
However, considering how the gauge field part will be in the classical Kerr solution case with a magnetic monopole~(e.g. see eq.(2.4) in \cite{Hartman:2008pb}), 
we would naturally reach the gauge field (\ref{RRBHTDG}).

Therefore, if we attempt to have a regular rotating black hole with a NLMM as a solution, 
because there remains no space where we can perform modification in the expression of the geometry with the gauge field, 
the way would be to add a new term to the Lagrangian as ${\cal L}+{\cal L}_{\rm Kerr}(a)$ only, where ${\cal L}_{\rm Kerr}(0)=0$. 
Considering to determine ${\cal L}_{\rm Kerr}(a)$ order-by-order, we expand as ${\cal L} + a^2{\cal L}^{(2)}_{\rm Kerr} + a^4{\cal L}^{(4)}_{\rm Kerr} + \cdots$, and consider ${\cal L}^{(2)}_{\rm Kerr}$ first. 
Then, it turns out that ${\cal L}^{(2)}_{\rm Kerr}$ fixed from the $(1,1)$, $(2,2)$, $(3,3)$ and $(4,4)$ components in the Einstein equation are respectively, 
\begin{eqnarray}
{\cal L}^{(2),\,(1,1)}_{\rm Kerr} = {\cal L}^{(2),\,(2,2)}_{\rm Kerr} \!\!&=&\!\! -\frac{24 m q_m^3 \cos ^2\theta \left(q_m^3+7 r^3\right)}{r^2\left(q_m^3+r^3\right)^3},\\
{\cal L}^{(2),\,(3,3)}_{\rm Kerr} = {\cal L}^{(2),\,(4,4)}_{\rm Kerr} \!\!&=&\!\!  \frac{1}{r^2} \left(C-\frac{120 m q_m^3 \cos ^2\theta}{\left(q_m^3+r^3\right)^2}\right),
\end{eqnarray}
where ${\cal L}^{(2),(i,i)}_{\rm Kerr}$~(i=1,2,3,4) stand for the ${\cal L}^{(2)}_{\rm Kerr}$ determined from these components in the Einstein equation, and $C$ is an integral constant with regard to the $r$-integral.  
Here the components other than these vanish at $a^2$-order, and do not give information for ${\cal L}^{(2)}_{\rm Kerr}$. 
Then, because ${\cal L}^{\textrm{(2),\,(1,1) or (2,2)}}_{\rm Kerr}$ and ${\cal L}^{\textrm{(2),\,(3,3) or (4,4)}}_{\rm Kerr}$ cannot agree with each other no matter how we take $C$, 
it can be seen that it is impossible to consider an action in which our regular rotating geometry with a NLMM can be an exact solution.

However, it turns out that in the non-zero components in the Einstein equation, $a$, $q_m$ and $r$ appear by the following order as
\begin{eqnarray}
\textrm{(1,1)}, \textrm{(2,2)} \!\!&\sim&\!\! a^2\left(\frac{q_m^3}{r^8}+\frac{q_m^6}{r^{11}}+\cdots\right)+a^4\left(\frac{q_m^3}{r^{10}}+\frac{q_m^6}{r^{13}}+\cdots\right)+\cdots=\sum_{i,j=1}^\infty a^{2j}   \frac{q_m^{3i}}{r^{3i+2j+3}},\nonumber \\ \\ 
\textrm{(1,4)}                 \!\!&\sim&\!\! a^3\left(\frac{q_m^3}{r^8}+\frac{q_m^6}{r^{11}}+\cdots\right)+a^5\left(\frac{q_m^3}{r^{10}}+\frac{q_m^6}{r^{13}}+\cdots\right)+\cdots=\sum_{i,j=1}^\infty a^{2j+1} \frac{q_m^{3i}}{r^{3i+2j+3}},\nonumber \\ \\
\textrm{(3,3)}, \textrm{(4,4)} \!\!&\sim&\!\! a^2\left(\frac{q_m^3}{r^6}+\frac{q_m^6}{r^9}+\cdots\right)+a^4\left(\frac{q_m^3}{r^8}+\frac{q_m^6}{r^{11}}+\cdots\right)+\cdots=\sum_{i,j=1}^\infty a^{2j} \frac{q_m^{3i}}{r^{3i+2j+1}},\nonumber \\
\end{eqnarray}
where numbers in the brackets above refer to the number of the component of the Einstein equation~(what is written are $[\rm{l.h.s}]-[\rm{r.h.s}]$ of it), and the above are the expressions in which coefficients are ignored.  
From this, we can consider that our regular rotating black hole geometry with a NLMM can be a solution approximately in the case of $q_m \ll r$. 
So, if $q_m \sim l_p$, where $l_p$ stands for the Plank length, the regular rotating geometry with a NLMM can be considered as a solution in the whole region except for the core region. 
In what follows, we treat $q_m$ as $l_p$.

\section{NHEK geometry in this study}
\label{Chap:NHEK} 

\subsection{$r_\pm$, $m$, $\tilde{\Delta}$ and $\Omega$ at the extremal limit}
\label{subChap:NHEK2} 

In this section, we obtain the NHEK geometry in our geometry with a NLMM given in section.\ref{Chap:RBH}. 
To this purpose, we first obtain $r_\pm$ and $m$ at the extremal limit, which can be obtained from $\tD = 0$
\footnote{
As the process leading to the extremality from non-extremal, normally the following three could be considered: 
i) With constant $a$, $m$ diminishes.\, 
ii) with constant $m$, $a$ grows.\, 
iii) both $m$ and $a$ vary.\. 
We consider the case (i). 
}. 
We denote those as $r_{\rm ext}$ and $m_{\rm ext}$. 
\newline

We can see that $\tD = 0$ has two real positive solutions, one real negative solution and two imaginary solutions. 
The two real positive solutions correspond to the outer and inner horizon radii. 
Rewriting as $\tD = 0 \rightarrow 2mr = (r^2+a^2)(1+l_p^3/r^3)$, 
at the moment that $2mr$ and $f(r) \equiv (r^2+a^2)(1+l_p^3/r^3)$ contact, the two real positive solutions become a multiple solution. 
This moment corresponds to the extremal limit, and this multiple solution corresponds to the horizon radius at the extremal limit. 
Let us consider the condition for what $2mr$ and $f(r)$ contact.

The tangential line of $f(r)$ can be written as $f'(r_{\rm ext})(r-r_{\rm ext})+f(r_{\rm ext})$.  
Therefore, a condition:~$2mr=f'(r_{\rm ext})(r-r_{\rm ext})+f(r_{\rm ext})$ holds, 
which leads to two conditions: $2m=f'(r_{\rm ext})$ and $r_{\rm ext} f'(r_{\rm ext})=f(r_{\rm ext})$.

It turns out that the equation to determine $r_{\rm ext}$ is a fifth-order equation.  
Therefore, considering to obtain $r_{\rm ext}$ up to the first correction of $l_p$, 
we once write the form of solution of $r_{\rm ext}$ as $r_{\rm ext}=(a^2+q_m^2)^{1/2} \left(1+\alpha \, l_p^3/(a^2+q_m^2)^{3/2} \right)$, where 
\begin{eqnarray}
l_p/(a^2+q_m^2)^{1/2} \ll 1. 
\end{eqnarray}
The analysis in what follows is performed under the expansion with regard to this $l_p/(a^2+q_m^2)^{1/2}$, 
and we omit to write ${\cal O}(l_p^6/(a^2+q_m^2)^3)$ for the simplicity of description.

Using the two conditions, $\alpha$ can be determined as $\alpha=3$.
Using this result, we can obtain $m_{\rm ext}$. 
Summarizing the results, our $r_{\rm ext}$ and $m_{\rm ext}$ are obtained as 
\begin{eqnarray} 
\label{raext} 
r_{\rm ext} \!\!&=&\!\! (a^2+q_m^2)^{1/2} \left(1      +  3 \frac{l_p^3}{(a^2+q_m^2)^{3/2}} \right), \\ 
\label{maext}
m_{\rm ext} \!\!&=&\!\! (a^2+q_m^2)^{1/2} \left(1      +   \frac{l_p^3}{(a^2+q_m^2)^{3/2}} \right).
\end{eqnarray}
Therefore, we can reach the extremality when $m_0$ reaches $m_{\rm ext}$ for a constant $a$.  
\newline

Next, let us obtain the expression of $\tD$ at the extremal limit. 
According to the statement at the beginning of this subsection, we can write as
\begin{eqnarray}\label{tD04}
\tD = \frac{1}{r^3 + l_p^3}(r-r_{\rm ext})^2(r-r_1)\big( (r-r_2)^2 + r_3^2 \big).
\end{eqnarray}
In what follows we obtain $r_1$, $r_2$ and $r_3$ up to the first correction of $l_p/(a^2+q_m^2)^{1/2}$.

Since it turns out again that the equation to determine $r_1$ is a fifth-order equation, 
writing $r_1$ as $r_1= -l_p \left( 1 + \beta l_p/(a^2+q_m^2)^{1/2} \right)$, we solve $[\tD \,\, \textrm{in eq.(\ref{tD})}]\big|_{r=r_1}=0$ with eq.(\ref{maext}).
As a result, $\beta$ can be determined as $\beta = -2/3$, which leads to
\begin{eqnarray}\label{r1} 
r_1=-l_p\left(1-\frac{2}{3} \frac{l_p}{(a^2+q_m^2)^{1/2}} \right).
\end{eqnarray}
Next, we obtain $r_2$ and $r_3$. 
Using the conditions obtained from [$\tD \,\, \textrm{in eq.(\ref{tD})}] = [\tD \,\, \textrm{in eq.(\ref{tD04})}$] with eqs.(\ref{raext}), (\ref{maext}) and (\ref{r1}), 
we can obtain $r_2$ and $r_3$ as
\begin{eqnarray}
r_2 \!\!&=&\!\! l_p \left( \frac{1}{2}        - \frac{1}{3} \frac{l_p}{(a^2+q_m^2)^{1/2}} \right), \\ 
r_3 \!\!&=&\!\! l_p \left( \frac{\sqrt{3}}{2} + \frac{1}{\sqrt{3}}\frac{l_p}{(a^2+q_m^2)^{1/2}} \right).
\end{eqnarray}

Therefore, up to the first correction of $l_p/(a^2+q_m^2)^{1/2}$,
\begin{eqnarray}\label{tDext}
\tD = \left(1 + \frac{2 l_p^3}{r^2(a^2+q_m^2)^{1/2}} \right)(r-r_{\rm ext})^2.
\end{eqnarray}
\newline

We here define the angular velocity $\Omega$ in this study. 
We adopt $\Omega$ obtained from the Killing vector field, $\xi=\partial_t+ \Omega \, \partial_\phi$, on the horizon.
In this case, at the extremal limit,
\begin{eqnarray}\label{AngleMomentaExt}
\Omega_{\rm ext} = \frac{a_{\rm ext}}{r_{\rm ext}^2+a_{\rm ext}^2}, 
\end{eqnarray}
where $\displaystyle \Omega_{\rm ext} \equiv \lim_{r \to r_{\rm ext}}\Omega$.

\subsection{Our NHEK geometry with the correction of the regularization effect}
\label{subChap:NHEK3} 

Now that we have obtained $r_{\rm ext}$, $m_{\rm ext}$ and other horizon radii at the extremal limit, 
let us obtain the NHEK geometry of eq.(\ref{RRBHTD}) with a gauge field (\ref{RRBHTDG}) in the co-rotating frame. 
The normal coordinates $(t, r, \theta, \phi)$ are related with the coordinate in the co-rotating frame $(\hat{t}, \hat{r}, \hat{\theta}, \hat{\phi})$ as 
\begin{eqnarray}\label{NH Limit}
\hat{t}    = \frac{\lambda \, t}{r_{\rm ext}}, \quad
\hat{r}    = \frac{r-r_{\rm ext}}{\lambda \, r_{\rm ext}}, \quad
\hat{\phi} = \phi - \frac{\Omega_{\rm ext} \,r_{\rm ext}\,t}{\lambda}
\quad \textrm{with} \quad \lambda \to 0, 
\end{eqnarray}
where taking the $\lambda \to 0$ limit corresponds to taking both extremal and near-horizon limits. 
The coordinates appearing in what follows are always the ones in the co-rotating frame, which we denote as $(t, r, \theta, \phi)$ without ~$\hat{}$~ in what follows.

Performing the following manipulation to the geometry (\ref{RRBHTD}) with a gauge field (\ref{RRBHTDG}) straightforwardly in the following sequence: 
\begin{enumerate}[label*=\arabic*)]
\item Substitute eq.(\ref{tDext}), 
\item substitute eq.(\ref{NH Limit}) with eq.(\ref{raext}), 
\item take the terms up to the $\l_p^3$ order, 
\item take the leading order in the $\lambda \to 0$ limit, 
\end{enumerate}
we can reach the NHEK geometry of eq.(\ref{RRBHTD}) as 
\begin{eqnarray}\label{NHEK}
ds^2
=
-   f_1 r^2 dt^2 
+   f_2 \frac{dr^2}{r^2}  
+   f_3 \, d\theta^2 
+   \Lambda^2      f_4 \, d\phi^2 
+ 2 \Lambda^2 \, r f_5 \, dt d\phi,
\end{eqnarray}
where $f_i = c_i + d_i \, l_p^3$ $(i=1,\cdots,5)$ with
\begin{eqnarray}\label{definition c}
(c_1,\,c_2,\,c_3,\,c_4,\,c_5) \!\!&=&\!\!
\Bigg(
\frac{\left(a^2 \cos ^2\theta+a^2+q_m^2\right)^2-4 a^2 \left(a^2+q_m^2\right) \sin ^2\theta}{a^2 \cos ^2\theta+a^2+q_m^2},
\nonumber\\
&& \hspace{-4mm}
a^2 \cos ^2\theta+a^2+q_m^2,\,\, a^2 \cos ^2\theta+a^2+q_m^2,
\nonumber\\
&& \hspace{-4mm}
\frac{\left(2 a^2+q_m^2\right)^2 \sin ^2\theta}{a^2 \cos ^2\theta+a^2+q_m^2}, \,\, \frac{2 a \sqrt{a^2+q_m^2} \left(2 a^2+q_m^2\right) \sin ^2\theta}{a^2 \cos ^2\theta+a^2+q_m^2}
\Bigg),
\end{eqnarray}
\begin{eqnarray}\label{definition d}
(d_1,\,d_2,\,d_3,\,d_4,\,d_5) 
\!&=&\!
\Bigg(
\frac{1}{16\left(a^2+q_m^2\right)^{3/2} \left(a^2 \cos ^2\theta+a^2+q_m^2\right)^2}
\bigg\{
\nonumber\\
&& \hspace{-32mm}
a^6 \cos (6 \theta )+306 a^6+696 a^4 q_m^2+528 a^2 q_m^4+128 q_m^6
\nonumber\\ 
&& \hspace{-32mm}
+\left(78 a^6+72 a^4 q_m^2\right) \cos (4 \theta )
+3a^2 \left(85 a^4+128 a^2 q_m^2+48 q_m^4\right) \cos (2 \theta )
\bigg\},
\nonumber\\
&& \hspace{-32mm} 
\frac{4 \left(a^2+q_m^2\right)-2 a^2 \cos ^2\theta}{\left(a^2+q_m^2\right)^{3/2}},\,\, 
\frac{6}{\sqrt{a^2+q_m^2}},\,\, 
\frac{24 \left(2 a^2+q_m^2\right) \sin ^2\theta \left(a^2 \cos (2 \theta )+a^2+q_m^2\right)}{\sqrt{a^2+q_m^2}\left(a^2 \cos (2 \theta )+3 a^2+2 q_m^2\right)^2},\nonumber\\
&& \hspace{-32mm} \frac{6 a \sin ^2\theta \left(a^2 \left(4 a^2+3 q_m^2\right) \cos ^2\theta+q_m^2\left(a^2+q_m^2\right)\right)}{\left(a^2+q_m^2\right) \left(a^2 \cos ^2\theta+a^2+q_m^2\right)^2}
\Bigg).
\end{eqnarray}
  
Now we write eq.(\ref{NHEK}) as 
\begin{eqnarray}
ds^2 = f_2\left\{ -\frac{\mu}{f_2}r^2dt^2+\frac{dr^2}{r^2}+\frac{f_3}{f_2}d\theta^2 + \frac{f_4}{f_2} \left( d\phi + \frac{f_5}{f_4}rdt \right)^2\right\},
\end{eqnarray}
where $\mu = f_1 + {f_5^2}/{f_4}$. 
Then defining  
\begin{eqnarray}
         c^2 \!\!&=&\!\! \frac{f_2}{\mu} = 1-\frac{4 l_p^3}{\left(a^2+q_m^2\right)^{3/2}}, \\
\bar{\kappa} \!\!&=&\!\! \frac{2 a \sqrt{a^2+q_m^2}}{2 a^2+q_m^2}-\frac{6 a q_m^2 l_p^3 }{\left(a^2+q_m^2\right) \left(2 a^2+q_m^2\right)^2},
\end{eqnarray}
we can write the above into
\begin{eqnarray}\label{NHEK0}
\frac{ds^2}{f_2} = - \frac{r^2}{c^2} dt^2 + \frac{dr^2}{r^2} + \frac{f_3}{f_2} d\theta^2 + \frac{f_4}{f_2} \left( d\phi + \bar{\kappa} \, r dt \right)^2,
\end{eqnarray}
where  
\begin{eqnarray}
\frac{f_3}{f_2} \!\!&=&\!\! \frac{2 a \sqrt{a^2+q_m^2}}{2 a^2+q_m^2}-\frac{6 a q_m^2 l_p^3}{\left(a^2+q_m^2\right)\left(2 a^2+q_m^2\right)^2},\\
\frac{f_4}{f_2} \!\!&=&  ~~~
\frac{\left(2 a^2+q_m^2\right)^2 \sin ^2\theta}{\left(a^2\cos ^2\theta+a^2+q_m^2\right)^2}\nonumber\\ 
&& +~ \frac{8 l_p^3 \left(2 a^2+q_m^2\right) \sin ^2\theta \big(a^2 \left(8 a^2+7 q_m^2\right) \cos (2\theta )+q_m^2 \left(a^2+2 q_m^2\right)\big)}
         {\left(a^2+q_m^2\right)^{3/2} \left(a^2 \cos(2 \theta )+3 a^2+2 q_m^2\right)^3}.
\end{eqnarray}
Then defining 
\begin{eqnarray}
r      \!\!&=&\!\! c \, \chi, \\
\kappa \!\!&=&\!\! c \, \bar{\kappa} = \frac{2 a \sqrt{a^2+q_m^2}}{2 a^2+q_m^2}-\frac{2 a l_p^3 \left(4 a^2+5 q_m^2\right)}{\left(a^2+q_m^2\right) \left(2 a^2+q_m^2\right)^2},
\end{eqnarray} 
it turns out that we can rewrite the NHEK (\ref{NHEK0}) into the following form:  
\begin{align}\label{NHEK1}
\frac{ds^2}{f_2} = -\chi^2dt^2+\frac{d\chi^2}{\chi^2}+\frac{f_3}{f_2}d\theta^2+\frac{f_4}{f_2}(d\phi+\kappa \chi dt )^2, 
\end{align}
where 
\begin{eqnarray}\label{NHEK1where}
\kappa \!\!&=&\!\!\! c \, k = \frac{2 a \sqrt{a^2+q_m^2}}{2 a^2+q_m^2}-\frac{2 a l_p^3 \left(4 a^2+5q_m^2\right)}{\left(a^2+q_m^2\right) \left(2 a^2+q_m^2\right)^2}. \quad
\end{eqnarray}
We can see from the form of the geometry (\ref{NHEK1}) that our NHEK geometry with the $l_p^3$ correction is also 
a $\theta$-dependent $S^1_{(\phi)}$-fibrated AdS$_2$ with an isometry SL(2,$\mathbb{R}$) $\times$ U(1)$_{(\phi)}$ as well as the NHEK geometry in \cite{Guica:2008mu}.

\section{Central charge in the dual CFT}
\label{Chap:CC} 

In this section, considering a set of the boundary conditions at the asymptotic region of the NHEK geometry obtained in the previous section, 
we compute the central charge in the isometries on this geometry. 
\newline

The boundary conditions $h_{\mu\nu}$ we set on the NHEK geometry $\bar{g}_{\mu\nu}$ in eq.(\ref{NHEK}) is the same with the one in \cite{Guica:2008mu} as
\begin{eqnarray}\label{deviation}
h_{\mu\nu}=
\left(
\begin{array}{cccc}
{\cal O}(r^2) & {\cal O}(1) & {\cal O}(1/r) & {\cal O}(1/r^2) \\
              & {\cal O}(1) & {\cal O}(1/r) & {\cal O}(1/r)   \\
              &             & {\cal O}(1/r) & {\cal O}(1/r^2) \\
              &             &               & {\cal O}(1/r^3) 
\end{array}
\right),
\end{eqnarray}
where $x_\mu=\left( t,\phi ,\theta,r\right)$. 
We write the metric of our NHEK geometry with the boundary conditions as 
\begin{eqnarray}\label{ghgh}
g_{\mu\nu}=\bar{g}_{\mu\nu} + h_{\mu\nu}. 
\end{eqnarray}

We now formally write the algebra of ASG~(asymptotic symmetry group) associated with this geometry at the semi-classical level as
\begin{align}\label{comLL}
[ L_m, L_n ] = (m-n) L_{m+n} + \frac{1}{\hbar}\frac{c_{\rm CFT}}{12} m(m^2-1) \, \delta_{m+n}, 
\end{align}
where $c_{\rm CFT}$ is the central charge, we treat $\hbar$ as $1$ in what follows. 
When the geometry given by $\bar{g}_{\mu\nu}$ can be written as
\begin{align}\label{NHEK2}
ds^2=\Gamma(\theta)\left\{-r^2 dt^2 + \frac{dr^2}{r^2} + \alpha(\theta)^2 d\theta^2 +\gamma(\theta)^2 (d\phi + \kappa \, r dt)^2\right\},  
\end{align} 
$c_{\rm CFT}$ in the case of the ASG associated with the boundary condition (\ref{deviation}) is known to be written as~\cite{Hartman:2008pb,Compere:2009dp}
\begin{align}\label{formulac}
c_{\rm CFT} = 3\kappa \int^\pi_0 \! d \theta \, \sqrt{\Gamma(\theta) \alpha(\theta) \gamma(\theta)}.  
\end{align}
Since the form (\ref{NHEK2}) is the same with eq.(\ref{NHEK1}), we can compute our central charge as  
\begin{align} \label{CCresult2}
3\kappa \int^\pi_0 \! d \theta \, \sqrt{\Gamma(\theta) \alpha(\theta)  \gamma(\theta)}\,\Big|_{\textrm{our NHEK (\ref{NHEK1})}} = 12 \left(a\sqrt{a^2+q_m^2}+\frac{a l_p^3}{a^2+q_m^2}\right).
\end{align}
We compute the same $c_{\rm CFT}$ in another way based on the Lagrangian formalism~\cite{Barnich:2001jy, Barnich:2007bf} in Appendix.\ref{App:CC}.

\section{Temperature in the dual CFT}
\label{Chap:CFTT} 

It is known~\cite{Guica:2008mu} that when the NHEK geometry is given by the form (\ref{NHEK1}), 
the dual CFT temperature is given by the Frolov-Thorne temperature $T_{\rm CFT}$~\cite{Frolov:1989jh} defined as
\begin{eqnarray}\label{Tphi1}  
T_{\rm CFT} \equiv - \lim_{T_H \to 0}\frac{T_H-0}{\Omega - \Omega_{\rm ext}} = -\frac{\partial T_H/\partial r}{\partial \Omega/\partial r}\bigg|_{r=r_{\rm ext}},
\end{eqnarray}
where $T_H$ is the Hawking temperature. 
In this section, we obtain $T_{\rm CFT}$ in our regular rotating black hole. 
\newline

We first write our original geometry (\ref{RRBHTD}) at the neighborhood of the extremal limit without the near-horizon limit in the following form:
\begin{eqnarray}\label{abstoriginalmetric}
ds^2 = g_{rr}dr^2 + g_{\theta\theta}d\theta^2 + a \left( a_t dt - a_\phi d\phi \right)^2 - b \left( b_t dt-b_\phi d\phi \right)^2,   
\end{eqnarray}
where  
\begin{eqnarray}
\label{TPgrrinv}
  g_{rr}^{-1} \!&=&\! \frac{\tD}{\Sigma} = \cG^{-1}(r- r_{\rm ext})^2, \\ 
\cG^{-1}      \!&=&\! \frac{1}{2} \partial_r^2 g_{rr}^{-1} \Big|_{r=r_{\rm ext}} 
                 =    \frac{1}{2} \partial_r^2 b \Big|_{r=r_{\rm ext}}  
                 =    \frac{1}{2} \partial_r^2\left( \frac{\tD}{\Sigma}\right) \Bigg|_{r=r_{\rm ext}} \nonumber \\ 
              \!&=&\!    \frac{1}{\Sigma} \frac{(r-r_1)\big((r-r_2)+r_3\big)^2}{r_{\rm ext}^3+l_p^3} \nonumber \\ 
              \!&=&\! \frac{1}{a^2 \cos ^2\theta+a^2+q_m^2}  +  \frac{l_p^3 \big(4 \left(a^2+q_m^2\right)-2 a^2 \cos ^2\theta\big)}{\left(a^2+q_m^2\right)^{3/2}}, \\ 
\textrm{and}\qquad\qquad\quad & & \nonumber \\
\label{TPgtt}
g_{\theta\theta} \!&=&\! \Sigma = r_{\rm ext}^2 + a^2 \cos \theta, \\
\label{TPa}
a \!&=&\! \frac{\sin^2 \theta}{\Sigma}, 
\quad
a_t = a, \,
\quad
a_\phi = r_{\rm ext}^2 + a^2, \\
\label{TPb}
b \!&=&\! g_{rr}^{-1} = \frac{\tD}{\Sigma} = \cG^{-1}(r- r_{\rm ext})^2, 
\quad             
\label{TPbt}
b_t = 1,
\quad
\label{TPbp}
b_\phi = a \sin^2 \theta.           
\end{eqnarray} 
We then rewrite the geometry (\ref{abstoriginalmetric}) into the co-rotating frame (\ref{NH Limit}) and take the $\lambda \to 0$ limit.  
At that time, each part in eq.(\ref{abstoriginalmetric}) can be written as
\begin{eqnarray}\label{ARGeo1}
                                       g_{rr} \, dr^2 \!\!&=&\!\! \cG \, \frac{dr^2}{r^2},\\
 a \, \left( a_t  \, dt - a_\phi  \, d \phi \right)^2 \!\!&=&\!\! a \, a_\phi{}^2 \left( \frac{r_{\rm ext}\,r}{\Omega_{\rm ext}} \partial_r \Omega \Big|_{r=r_{\rm ext}} \, dt - d\phi \right)^2,\\
-b \, \left( b_t  \, dt - b_\phi  \, d \phi \right)^2 \!\!&=&\!\! - \cG^{-1} \, r_{\rm ext}{}^2\,r^2 \left( \frac{b_t}{\Omega_{\rm ext}} - b_\phi \right)^2  dt^2.
\end{eqnarray}
Performing the following transformation:
\begin{eqnarray} 
t = c\,\tau, \quad \textrm{where $c$ satisfies} \quad\! c^2 \, \frac{r_{\rm ext}{}^2}{\cG^2} \left( \frac{b_t}{\Omega_{\rm ext}} - b_\phi \right)^2 = 1,  
\end{eqnarray}  
eq.(\ref{abstoriginalmetric}) can be written as
\begin{eqnarray}\label{61b}
\frac{ds^2}{\cG} = \left( \frac{dr^2}{r^2} - r^2 d \tau^2 \right) 
                   + \frac{g_{\theta\theta}}{\cG} \, d\theta^2
                   + \frac{a \, a_\phi^2}{\cG} \left( d\phi - \frac{\partial_r \Omega \big|_{r=r_H}}{\frac{1}{2}(b_t - \Omega_{\rm ext} b_\phi) \, \partial_r^2 b \big|_{r=r_H}} \, r d\tau \right)^2,
\end{eqnarray}
where we can confirm the following agreements: 
\begin{eqnarray}
\label{km1}
{\cal G}                      \,=\, f_2, \quad
\frac{g_{\theta\theta}}{\cG}  \!\!&=&\!\! \frac{f_3}{f_2}, \quad
\frac{a \, a_\phi^2}{\cG}        \,=\,  \frac{f_4}{f_2}, \nonumber\\
\label{km} 
\frac{\partial_r \Omega \big|_{r=r_H}}{\frac{1}{2} \, (b_t - \Omega_{\rm ext} b_\phi) \, \partial_r^2 b \big|_{r=r_H}} \!\!&=&\!\! 
-\frac{2 a \sqrt{a^2+q_m^2}}{2a^2+q_m^2} + \frac{2 l_p^3 \left(4 a^3+5 a q_m^2\right)}{\left(a^2+q_m^2\right)\left(2 a^2+q_m^2\right)^2} = -\kappa.
\end{eqnarray}
Here, $f_2$,\, ${f_3}/{f_2}$,\, ${f_4}/{f_2}$ and $\kappa$ mean the quantities in eq.(\ref{NHEK1}).  
On the other hand, since the Hawking temperature $T_H$ can be represented as in (\ref{THRepr}),
\begin{eqnarray}\label{61c}
\textrm{Eq.(\ref{km1})}
=\frac{\partial_r \Omega \big|_{r=r_H}}{\frac{1}{2} \, (b_t - \Omega_{\rm ext} b_\phi) \, \partial_r^2 b \big|_{r=r_H}}
=\frac{1}{2\pi}\frac{\partial_r \Omega \big|_{r=r_H}}{\partial_r T_H\big|_{r=r_H}}
=-\frac{1}{2\pi}\frac{1}{T_{\rm CFT}}.
\end{eqnarray}
Therefore, we can obtain $T_{\rm CFT}$ as 
\begin{eqnarray}\label{CFTTempresult}
T_{\rm CFT} = \frac{2 a^2+q_m^2}{4 \pi  a\sqrt{a^2+q_m^2}} + \frac{l_p^3 \left(4 a^2+5 q_m^2\right)}{4 \pi  a\left(a^2+q_m^2\right)^2}. 
\end{eqnarray}

\section{Entropy in the dual CFT}
\label{Chap:EdCFT} 

We now compute the entropy in the dual CFT using the Cardy formula: 
\begin{eqnarray}\label{cardyf}
S_{\rm CFT}=\frac{\pi^2}{3} c_{\rm CFT} \, T_{\rm CFT}. 
\end{eqnarray}
$c_{\rm CFT}$ and $T_{\rm CFT}$ have been obtained in eqs.(\ref{CCresult2}) and (\ref{CFTTempresult}), and we can compute $S_{\rm CFT}$ as
\begin{eqnarray}\label{CFTEntropy}
S_{\rm CFT} = \pi  \left(2 a^2+q_m^2\right) + \frac{6 \pi  l_p^3}{\sqrt{a^2+q_m^2}}.
\end{eqnarray} 
We also evaluate \vspace{1mm} the Bekenstein-Hawking entropy, $S_{\rm BH}={\cal A}/{4}$ where $\displaystyle {\cal A}=\int_0^{2\pi} \!d\phi \int_0^{\pi} \!d\theta \sqrt{g_{\theta\theta}\,g_{\phi\phi}}$, 
in our geometry at the extremal limit. 
We can confirm that it can agree with $S_{\rm CFT}$.

\section{Summary}
\label{Chap:Summary}

In this study, considering a rotating Hayward type regular black hole with a NLMM, we have carried out the Kerr/CFT correspondence at the extremal limit.

One problem in this study would have been whether our geometry can be a solution or not.  
In this study, based on \cite{Fan:2016hvf} which prescribes the Einstein-nonlinear Maxwell models 
which contains the regular Schwarzschild black hole geometry with a non-linear electron or magnetic monopole as a solution,  
we have searched for the way to also make our rotating geometry with a NLMM into a solution. 
Then it has turned out that it is adding a new term vanishing at zero-rotation to the action, because the form of the geometry seems to have no space where we can add further modification for making into a solution.   
This would be the natural extension we can consider from the Schwarzschild case~\cite{Fan:2016hvf}. 
However, it has turned that we cannot obtain the action exactly containing the rotation effect of our geometry as a solution. 
This might indicate that since the extension to the Kerr case has been impossible, there might be some problem in the way~\cite{Fan:2016hvf} in a sense that it is succeeded only in the Schwarzschild case. 
In this study, we have managed this problem by using the fact that our geometry can be a solution approximately in the whole space-time except for the core region of the black hole.

Next problem in this study has been that the equation to obtain the horizon radii is a fifth-order equation due to the regularization effect.   
In order to overcome this, we have treated the regularization effect perturbatively.  
As a result, we could obtain the NHEK geometry with the correction of the regularization effect.

Once we could obtain the NHEK geometry, we could obtain the central charge and read off the CFT temperature.  
Then using these, we could compute the entropy in the dual CFT which agrees with the Bekenstein-Hawking entropy.

\section*{Acknowledgment}
Seiji Terashima could kindly check some calculation in this paper with author and help find out miscalculation. 
Author would like to express appreciation to him so much. 
Author would also be very grateful for Kiyoshi Sogo's very helpful advice on how to obtain the horizon radii at the extremal limit in Sec.\ref{subChap:NHEK2}. 
\appendix 

\section{Computation of the central charge}
\label{App:CC}

In this appendix, we compute $c_{\rm CFT}$ based on the Lagrangian formalism~\cite{Barnich:2001jy, Barnich:2007bf}, the result of which agrees with $c_{\rm CFT}$ (\ref{CCresult2}).

\subsection{Asymptotic Killing vector preserving the ASG}
\label{AppSubChap:killing vec} 

In the geometry (\ref{ghgh}), the most general Killing vector belonging to the allowed diffeomorphisms\footnote{
\begin{eqnarray}
\textrm{ASG}=\frac{\textrm{Group of allowed diffeomorphisms}}{\textrm{Group of trivial diffeomorphisms}}.
\end{eqnarray}
`Group of allowed diffeomorphisms' means the group of the diffeomorphisms generated by the most general Killing vectors $\xi=\xi^\mu\partial_\mu$.   
On the other hand, `Group of trivial diffeomorphisms' means a subgroup of `Group of allowed diffeomorphisms' 
consisted of the Killing vectors with no contribution to the conserved charge.} can be written as~\cite{Guica:2008mu}
\begin{eqnarray}\label{Killing vector1}
\xi =  
  \left( C + {\cal O}(r^{-3})            \right) \partial_t 
+ \left( -r \epsilon'(\phi)+{\cal O}(1) \right) \partial_r 
+ \left( \epsilon(\phi)+{\cal O}(r^{-2}) \right) \partial_\phi
+ \left( {\cal O}(r^{-1})                  \right) \partial_\theta,
\end{eqnarray}
where $C$ is some constant and $\epsilon(\phi)$ is some function of $\phi$. 
However, since the $\phi$-direction in the NHEK geometry is periodic by $2\pi$, $\epsilon(\phi)$ is periodic as $\epsilon(\phi)=\epsilon(\phi+2\pi)$.

The generic form of the conserved charges associated with the diffemorphisms can be written as $Q = \int_{\rm boundary} X + \int_{\rm bulk} Y$, 
where $X$ and $Y$ are some conserved currents. 
The `bulk' term  typically vanish due to the constraint equations. 
Hence, only the `boundary' term can give the contribution to the conserved charges (and diffemorphisms are generated by the `boundary' term).   
Therefore, the subleading terms in eq.(\ref{Killing vector1}) correspond to the trivial diffeomorphisms, 
because these decay in the bulk and cannot reach the boundary,
and make no contribution to the conserved charges.

In addition, writing the conserved charges associated with the $t$-translation as $Q_t$, the values of $Q_t$ correspond to a deviation from the energy at the extremal limit. 
We study the extremal situation in which~$Q_t=0$. 
Hence, the leading term $C$ in the $\partial_t$-term  in eq.(\ref{Killing vector1}) belongs to the trivial diffeomorphism.

Therefore, the asymptotic Killing vector part in eq.(\ref{Killing vector1}) is
\begin{eqnarray}
\xi = \epsilon(\phi) \partial_\phi - r \epsilon'(\phi) \partial_r.  
\end{eqnarray} 
In what follows, we consider this $\xi$. 
As mentioned above, $\epsilon(\phi)$ is periodic, as well as \cite{Guica:2008mu}, we expand $\epsilon(\phi)$ as 
$\displaystyle 
\epsilon(\phi) = -\sum_{n=-\infty}^{\infty} e^{-in\phi} 
$. 
Then, $\xi$ can be written as 
\begin{eqnarray}\label{xin}
        \xi =       -\sum_{n=-\infty}^{\infty} e^{-in\phi} (\partial_\phi + inr\partial_r)  
 \,\,\,     =   \sum_{n=-\infty}^{\infty} e^{-in\phi} \xi_n, 
\quad {\rm where} \quad \xi_n \equiv -(\partial_\phi + inr \partial_r),  
\end{eqnarray}
and we can find that the Lie bracket of $\xi_n$ forms the one copy of the Virasoro algebra as
\begin{eqnarray}\label{LBxixi}
i[\xi_m,\xi_n]_{\rm L}=(m-n)\xi_{m+n}.
\end{eqnarray}

\subsection{Central charge}
\label{AppSubChap:CC}  

Writing the generators of the diffeomorphisms generated by $\xi$ as $\mathbf{H}[\xi]=H[\xi]+Q[\xi]$, 
the generic form of the Poisson bracket of $\mathbf{H}[\xi]$ can be written as
\begin{eqnarray}\label{HH}
\big\{ \mathbf{H}[\xi],\mathbf{H}[\eta] \big\}_{\rm{P}} = \mathbf{H}\big[[\xi,\eta]_{\rm L}\big] + K\big[\xi,\eta\big],
\end{eqnarray}
where $H[\xi]$ and $Q[\xi]$ are respectively the bulk and boundary parts, and $K[\xi,\eta]$ stands for the central charge. 
Since we can suppose that the system we are treating now has the first class constraints, we impose a gauge fixing conditions, and introduce the Dirac bracket, 
which satisfies the same algebra with the Poisson bracket, but $H[\xi]$ vanishes in the Dirac bracket. Hence from eq.(\ref{HH}), we can write as 
\begin{eqnarray}\label{Virasolo algebra1}
\big\{ Q[\xi],Q[\eta] \big\}_{\rm D} = Q\big[[\xi,\eta]_{\rm L}\big] + K\big[\xi,\eta\big]=\delta_\eta Q\big[\xi\big],    
\end{eqnarray}
where $\delta_\eta$ means the difference in the charge's value when a given space-time is displaced by $\eta$. 
From the above, 
we can write the central charge as
\begin{eqnarray}
K[\xi,\eta] = \delta_\eta Q[\xi] \big|_{g_{\mu\nu}=\bar{g}_{\mu\nu}},  
\end{eqnarray}
where the constant part in $Q\big[ \xi \big]$ is taken so that $Q\big[\xi\big] \big|_{ g_{\mu\nu} = \bar{g}_{\mu\nu}}=0$ for general $\xi$,   
and $\eta$ in the above can be now regarded as $h_{\mu\nu}$ in eq.(\ref{deviation}).

It is known that $\delta_\eta Q[\xi] \big|_{g_{\mu\nu}=\bar{g}_{\mu\nu}}$ can be computed as~\cite{Barnich:2001jy} 
\begin{eqnarray}\label{dQ1}
\delta_\eta Q[\xi] \big|_{g_{\mu\nu}=\bar{g}_{\mu\nu}} = \frac{1}{8\pi G}\int_{\partial \Sigma}k_{\xi}[{\cal L}_n\bar{g},\bar{g}],
\end{eqnarray} 
where $G=1$ in the following, ${\cal L}_n \equiv {\cal L}_{\xi_n}$ are Lie derivatives, $\partial \Sigma$ is the boundary of the spatial surface. 
Further, since $h_{\mu\nu}$ have the same $r$-dependence with ${\cal L}_n\bar{g}_{\mu\nu}$, we have regarded as 
$\eta_{\mu\nu}=h_{\mu\nu}={\cal L}_n\bar{g}_{\mu\nu}$, and
\begin{eqnarray}\label{kxi1}
k_{\xi}[h,\bar{g}]   
\!\!\!&=&\!\!\! 
-\frac{\sqrt{-\det \bar{g}}}{4}\,\epsilon_{\alpha \beta \mu \nu} \,
\bigg\{
  \xi^\nu    D^\mu    h 
- \xi^\nu    D_\sigma h^{\mu\sigma} 
+ \xi_\sigma D^\nu    h^{\mu\sigma} 
-h^{\nu\sigma}D_\sigma \xi^\mu 
\nonumber \\ && \qquad\qquad\qquad\qquad \!
+ \frac{1}{2} h D^\nu \xi^\mu 
+ \frac{1}{2} h^{\sigma\nu} \Big( D^\mu \xi_\sigma - D_\sigma \xi^\mu \Big)
\bigg\} \, dx^\alpha \wedge dx^\beta,  
\end{eqnarray}
where the covariant derivatives are given by $\bar{g}_{\mu\nu}$, raising/lowering of the indices are performed using $\bar{g}_{\mu\nu}$, and 
$h = \bar{g}^{\mu\nu}h_{\mu\nu}$, and $\epsilon^{t\theta\phi r}=1$. 
Therefore from eq.(\ref{Virasolo algebra1}), 
\begin{eqnarray}\label{dQ2}
\{ Q_{m}, Q_{n} \}_{\rm D} 
\!\!&=&\!\! Q_{[\xi_m,\xi_n]_{\rm L}} + \frac{1}{8\pi} \int_{\partial\Sigma} k_{\xi_m}[{\cal L}_n\bar{g},\bar{g}] \nonumber\\  
\!\!&=&\!\! -i(m-n) Q_{m+n} + \frac{1}{8\pi} \int_{\partial\Sigma} k_{\zeta_m}[{\cal L}_n\bar{g},\bar{g}], 
\end{eqnarray} 
where $Q_{m}\equiv Q_{\xi_{m}}$, and $Q_{[\zeta_m,\zeta_n]_{\rm L}}=-i(m-n) Q_{m+n}$.

\vspace{+2.0mm}
In our geometry (\ref{NHEK1}), $k_{\xi}[h,\bar{g}]$ can be computed as
\begin{eqnarray}\label{kxi2}
k_{\xi}[h,\bar{g}]  
\!\!\!&=&\!\!\!
\frac{\sqrt{-\det \bar{g}}}{2} \,
\bigg\{
  \xi^r \bar{g}^{t\alpha}\bar{g}_{\mu\nu}D_\alpha h^{\mu\nu} 
- \xi^r D_\sigma h^{t \sigma} 
+ \bar{g}_{\sigma\beta} \xi^\beta \bar{g}^{r\alpha} D_\alpha h^{t \sigma} 
+ \frac{1}{2} h \bar{g}^{r\alpha} D_\alpha \xi^t
\nonumber \\ && \qquad
\!\!\!  
- h^{r \sigma} D_\sigma \xi^t 
+ \frac{1}{2}h^{\sigma r}
\Big(
\bar{g}^{t\beta} \bar{g}_{\sigma\alpha} D_\beta \xi^\alpha + D_\sigma \xi^t
\Big)
\nonumber \\ && \qquad
\!\!\!
- \bar{g}_{\sigma\beta} \xi^\beta \bar{g}^{t\alpha} D_\alpha h^{r \sigma}
- \frac{1}{2} h \bar{g}^{t\alpha} D_\alpha \xi^r
+ \frac{1}{2} h^{t\sigma} D_\sigma \xi^r 
- \frac{1}{2} h^{\sigma t} \bar{g}^{r\beta} \bar{g}_{\sigma\alpha} D_\beta \xi^\alpha 
\bigg\} \, d\theta \wedge d\phi \nonumber \\ && 
+ \cdots \nonumber \\
&=&
\Big\{
 {\cal J}_1 h_{tt}\epsilon
+{\cal J}_2 h_{\phi\phi}\epsilon
+{\cal J}_3 (h_{r\phi}\epsilon'-h_{r\phi}' \epsilon)
\Big\} \,d\theta \wedge d\phi
\nonumber\\
&&
+
\, l_p^3 \,
\Big\{
  \epsilon \, {\cal K}_1 h_{tt} 
+ \epsilon \, {\cal K}_2 h_{\phi\phi}
+ \, {\cal K}_3 
\Big(
\epsilon' h_{r\phi} - \epsilon h_{r\phi}' 
\Big) 
\Big\} 
\,d\theta \wedge d\phi,
\end{eqnarray}  
where $'$ means $\partial_\phi$, ``$+\cdots$'' in the first line means the terms not to contribute in the integration which we disregard in the second line, and 
\begin{eqnarray}
{\cal J}_1 \!\!&=&\!\! -\frac{a \sqrt{a^2+q_m^2} \left(2 a^2+q_m^2\right)^2 \sin ^3\theta}{2 \left(a^2 \cos ^2\theta+a^2+q_m^2\right)^3}, \\
{\cal J}_2 \!\!&=&\!\! -\frac{a \sqrt{a^2+q_m^2} \sin \theta \left(a^4 \cos (4 \theta )+35 a^4-4 a^2 \left(a^2+2 q_m^2\right) \cos (2 \theta )+40 a^2 q_m^2+8 q_m^4\right)}{16 \left(a^2 \cos ^2\theta+a^2+q_m^2\right)^3}, \nonumber\\ \\
{\cal J}_3 \!\!&=&\!\! \frac{a \sqrt{a^2+q_m^2} \sin \theta}{a^2 \cos ^2\theta+a^2+q_m^2}, \\
{\cal K}_1 \!\!&=&\!\! 
-\frac{4 a \left(2 a^2+q_m^2\right) \sin ^3\theta \left(-30 a^4+a^2 \left(14 a^2+13 q_m^2\right) \cos (2 \theta )-41 a^2 q_m^2-10 q_m^4\right)}
{\left(a^2+q_m^2\right) \left(a^2 \cos (2 \theta )+3 a^2+2q_m^2\right)^4}, \\
{\cal K}_2 \!\!&=&\!\! 
-\frac{1}{4 \left(a^2+q_m^2\right) \left(a^2 \cos (2 \theta )+3 a^2+2 q_m^2\right)^4}
\bigg\{
\nonumber\\
&&
a \sin \theta 
\bigg(
3 a^6 \cos (6 \theta ) -670 a^6-1644 a^4 q_m^2-1040 a^2 q_m^4-96q_m^6\nonumber\\
&&
-2 a^4 \left(41 a^2+50 q_m^2\right) \cos (4 \theta )
+a^2 \left(749 a^4+1360 a^2 q_m^2+656 q_m^4\right) \cos (2 \theta ) 
\bigg)
\bigg\}
, \\
{\cal K}_3 \!\!&=&\!\! \frac{6 a \sin \theta \left(a^2 \cos (2 \theta )-a^2-2 q_m^2\right)}{\left(a^2+q_m^2\right) \left(a^2 \cos (2 \theta )+3 a^2+2 q_m^2\right)^2}.
\end{eqnarray}
As 
$ h_{\mu\nu}= {\cal L}_n g_{\mu\nu} = \xi^\lambda \partial_\lambda g_{\mu\nu} + g_{\lambda\nu} \partial_\mu \xi^{\lambda} + g_{\mu\lambda} \partial_\nu \xi^{\lambda}$, 
performing the computation for each $h_{\mu\nu}$ appearing in eq.(\ref{kxi2}) as
\begin{eqnarray}
{\cal L}_n g_{tt}       \!\!&=&\!\! 2 \, r^2 \left( {\cal J}_1 + {\cal K}_1 l_p^3 \right) \, ine^{-in\phi},\\
{\cal L}_n g_{r\phi}    \!\!&=&\!\! - r^{-1} \left( {\cal J}_2 + {\cal K}_2 l_p^3 \right) n^2 e^{-in\phi},\\
{\cal L}_n g_{\phi\phi} \!\!&=&\!\! 2 \left( {\cal J}_3 + {\cal K}_3 l_p^3 \right) e^{-in\phi},
\end{eqnarray} 
we can compute the central charge (\ref{dQ2}) as 
\begin{eqnarray}\label{calcCC1} 
\frac{1}{8\pi} \int_{\partial \Sigma} k_{\xi}[{\cal L}_n\bar{g},\bar{g}] 
\!\! &=& \!\! 
i \int_0^{2\pi} \! d\phi \int_0^{\pi} \! d\theta \, e^{-i(m+n)\phi}\, 
\Bigg[ 
\nonumber \\ && 
\frac{a n \sqrt{a^2+q_m^2} \sin \theta \left(n (n-m) \left(a^2 \cos ^2\theta+a^2+q_m^2\right)^2+2 \left(2a^2+q_m^2\right)^2 \sin ^2\theta\right)}{\left(a^2 \cos ^2\theta+a^2+q_m^2\right)^2}
\nonumber \\ && 
\frac
{a l_p^3 n}{8 \left(a^2+q_m^2\right) \left(a^2 \cos (2 \theta )+3 a^2+2q_m^2\right)^3}
\bigg\{
a^6 n \sin (7 \theta ) (n-m)
\nonumber \\ && 
-a^2 \sin (5 \theta ) 
\Big(a^4 (17 n (m-n)+288)+12 a^2 q_m^2 (n (m-n)+32)+120q_m^4\Big)
\nonumber \\ && 
+\sin \theta 
\Big(
-3 a^6 (47 n (m-n)+192)-24 a^4 q_m^2 (13 n (m-n)+16)
\nonumber \\ && \qquad\qquad \!\!
+\, 48 a^2 q_m^4 (5 n(n-m)+11)+32 q_m^6 (2 n (n-m)+9)
\Big)
\nonumber \\ && 
+3 \sin (3 \theta ) 
\Big(
~~ a^6 (31 n (n-m)+224)+4 a^4 q_m^2 (11 n (n-m)+64)
\nonumber \\ && \qquad\qquad\quad
+8 a^2q_m^4 (2 n (n-m)+1)-32 q_m^6
\Big)
\bigg\}
\Bigg]
\nonumber\\
\!\! &=& \!\!
-im 
\Bigg[ \! 
a m^2 \sqrt{a^2+q_m^2} \nonumber\\ && ~~~~~
- \frac{\left(2 a^2 +q_m^2\right)^2}{4a^2\left(a^2+q_m^2\right)}
\left\{
q_m^2 \tan ^{-1}\left(\frac{2 a\sqrt{a^2+q_m^2}}{q_m^2}\right) - 2a \sqrt{a^2+q_m^2}
\right\}
\nonumber \\ && ~~~~~
+l_p^3 \Bigg\{
\frac{a m^2}{a^2+q_m^2}
+\frac{1}{2 a^2 \left(a^2+q_m^2\right)^3}
\Bigg(
12 a^7+42 a^5 q_m^2+42 a^3 q_m^4+12a q_m^6
\nonumber \\ && \qquad \quad~~~
-3 \sqrt{a^2+q_m^2} \left(4a^6+8 a^4 q_m^2+7 a^2 q_m^4+2 q_m^6\right) \nonumber \\ && \qquad \quad~~~
\times \tan ^{-1}\left(\frac{2 a \sqrt{a^2+q_m^2}}{q_m^2}\right)
\Bigg)
\Bigg\} \Bigg] \, \delta_{m+n,0}, 
\nonumber\\
&\equiv& -i m\left\{ p_0 \left(\frac{c_0}{p_0}+m^2\right)+l_p^3 p_3 \left(\frac{c_3}{p_3}+m^2\right)\right\},    
\end{eqnarray} 
where we have used $\displaystyle \int_0^{2\pi} \! d\phi \, e^{-i(m+n)\phi}=2\pi \, \delta_{m+n,0}$, and 
\begin{eqnarray}
(p_0,\,\, p_3) \!\!&=&\!\! \left(a \sqrt{a^2+q_m^2},\,\, \frac{a}{a^2+q_m^2} \right), \\
(c_0,\,\, c_3) \!\!&=&\!\! 
\Bigg(
-\frac{\left(2 a^2 +q_m^2\right)^2}{4a^2\left(a^2+q_m^2\right)}
\left\{
q_m^2 \tan ^{-1}\left(\frac{2 a\sqrt{a^2+q_m^2}}{q_m^2}\right) - 2a \sqrt{a^2+q_m^2}
\right\},
\nonumber \\ && \hspace{-13mm}
\frac{1}{2 a^2 \left(a^2+q_m^2\right)^3} 
\Bigg\{
12 a^7+42 a^5 q_m^2+42 a^3 q_m^4+12a q_m^6
\nonumber \\ && \hspace{-13mm}
-3 \sqrt{a^2+q_m^2} \left(4a^6+8 a^4 q_m^2+7 a^2 q_m^4+2 q_m^6\right) \tan ^{-1}\left(\frac{2 a \sqrt{a^2+q_m^2}}{q_m^2}\right)
\Bigg\}
\Bigg).
\end{eqnarray} 
Hence we can write eq.(\ref{dQ2}) as
\begin{eqnarray}\label{QQD}
\{ Q_{m}, Q_{n} \}_{\rm D} = -i(m-n) Q_{m+n} -i m\left\{ p_0 \left(\frac{c_0}{p_0}+m^2\right)+l_p^3 p_3 \left(\frac{c_3}{p_3}+m^2\right)\right\} \, \delta_{m+n,0}. 
\end{eqnarray}

In order to obtain the quantum theory version of eq.(\ref{QQD}), we perform the replacement:~$\{., .\}_{\rm D} \rightarrow i\,[., .]$ with 
\begin{eqnarray}\label{replaceQn}
\displaystyle Q_{n} \rightarrow \hbar L_n - \frac{1}{2}\left(q_0+ q_3\,l_p^3\right) \delta_{n,0}.
\end{eqnarray} 
The coefficients $q_0$ and $q_3$ are determined so that 
the algebra obtained by the above replacement takes the form of the Virasolo algebra with a central charge $ c_{\rm CFT}$ in 2D CFT such as 
$\displaystyle [ L_m, L_n ] = (m-n) L_{m+n} + \frac{J \, c_{\rm CFT}}{12} m(m^2-1) \delta_{m+n}$, 
in other words, so that the constant part to be the central term can be bundled with $m(m^2-1)$. 
It turns out that $q_0$ and $q_3$ should be taken as 
\begin{eqnarray} \label{q0q3}
\frac{q_i-c_i}{p_i} = 1, \qquad i=0,\,3. 
\end{eqnarray}
After all, the quantum theory version of eq.(\ref{QQD}) can be obtained as
\begin{align}\label{comLL}
[ L_m, L_n ] = (m-n) L_{m+n} + \left(p_0+l_p^3 p_3\right) m(m^2-1) \, \delta_{m+n,0}. 
\end{align} 
From this, we can read off the central charge in the ASG in our NHEK geometry as 
\begin{eqnarray} \label{CCresult1}
c_{\rm CFT} = 12 \left( a\sqrt{a^2+q_m^2} + \frac{a l_p^3}{a^2+q_m^2} \right). 
\end{eqnarray}

\section{Expression of the Hawking temperature in our geometry}
\label{App:CFTT}

The Hawking temperature $\displaystyle T_H$ at the extremal limit in our model can be written as 
\begin{eqnarray}\label{THRepr}
 T_H \!\!&=&\!\! \frac{1}{2\pi}\sqrt{-\frac{1}{2}g^{ac}g^{bd}\nabla_a \xi_b \nabla_c \xi_d}\Bigg|_{r={\rm ext}} \nonumber \\
     \!\!&=&\!\! \frac{1}{2\pi}\Big[ -\frac{1}{2} g^{rr} \Big( g^{tt}(\nabla_r \xi_t)^2 + g^{\phi\phi}(\nabla_r \xi_\phi)^2 + 2 g_{t\phi} \nabla_r \xi_t \nabla_r \xi_\phi \Big) \nonumber \\
\label{THGeometryU}
     & & \,\qquad - \frac{1}{2}g^{rr} \Big( g^{tt}(\nabla_t \xi_r)^2 + g^{\phi\phi}(\nabla_\phi \xi_r)^2 + 2 g_{t\phi} \nabla_t \xi_r \nabla_\phi \xi_r \Big) \Big]^{1/2}\bigg|_{r={\rm ext}} \nonumber\\
\label{THGeometry}
     \!\!&=&\!\! \frac{1}{4\pi} \left( b_t - b_\phi \, \Omega_{\rm ext} \right) \partial_r b \Big|_{r =r_{\rm ext}},
\end{eqnarray}
where 
\begin{eqnarray}\label{Omegaii}
\xi = \xi^\mu \partial_\mu = \partial_t + \Omega \, \partial_\phi \quad{\rm with}\quad \Omega = \frac{a}{a^2+r^2}, 
\end{eqnarray}
and
\begin{eqnarray}  
\label{Dxi_in_TH1}
\nabla_r \xi_t    \Big|_{r =r_{\rm ext}} \!\!&=&\!\! -\frac{a \left\{\Omega \left(2 r+\Delta '\right)+2 \Omega ' \left(a^2+r^2\right) \right\}\sin ^2\theta+\Delta '}{2 \Sigma}, \\
\label{Dxi_in_TH2}
\nabla_r \xi_\phi \Big|_{r =r_{\rm ext}} \!\!&=&\!\!  \frac{\sin ^2\theta}{2 \Sigma} \left[\Omega \left\{4 r \left(a^2+r^2\right)-a^2 \Delta ' \sin ^2\theta  \right\}+2 \Omega ' \left(a^2+r^2\right)^2 +a \Delta '-2 a r\right], \nonumber\\ \\
\label{Dxi_in_TH3}
\nabla_t \xi_r    \Big|_{r =r_{\rm ext}} \!\!&=&\!\!  \frac{a \Omega \left(2 r-\Delta '\right) \sin ^2\theta  +\Delta '}{2 \Sigma}, \\
\label{Dxi_in_TH4}
\nabla_\phi \xi_r \Big|_{r =r_{\rm ext}} \!\!&=&\!\!  \frac{\sin ^2\theta}{2 \Sigma} \left[2 r \left\{a-2 \Omega\left(a^2+r^2\right) \right\}+a \Delta ' \left(a \Omega \sin ^2\theta  -1\right)\right].
\end{eqnarray}

\end{document}